\documentclass[aps,prb,twocolumn,superscriptaddress]{revtex4-1}

\usepackage{graphicx}
\usepackage{dcolumn}
\usepackage{bm}
\usepackage{amsmath}
\usepackage{todonotes}

\begin{document}


\title{Spin-dependent recombination involving oxygen-vacancy complexes in silicon}


\author{David P. Franke}
\email[]{david.franke@wsi.tum.de}
\author{Felix Hoehne}
\affiliation{Walter Schottky Institut, Technische Universit\"at M\"unchen, Am Coulombwall 4, 85748 Garching, Germany}
\author{Leonid S. Vlasenko}
\affiliation{A. F. Ioffe Physico-Technical Institute, Russian Academy of Sciences, 194021, St.~Petersburg, Russia}
\author{Kohei M. Itoh}
\affiliation{School of Fundamental Science and Technology, Keio University, 3-14-1 Hiyoshi, Kohuku-ku, Yokohama 223-8522, Japan}
\author{Martin S. Brandt}
\affiliation{Walter Schottky Institut, Technische Universit\"at M\"unchen, Am Coulombwall 4, 85748 Garching, Germany}

\date{\today}

\begin{abstract}
Spin-dependent relaxation and recombination processes in $\gamma$-irradiated $n$-type Czochralski-grown silicon are studied using continuous wave (cw) and pulsed electrically detected magnetic resonance (EDMR). Two processes involving the SL1 center, the neutral excited triplet state of the oxygen-vacancy complex, are observed which can be separated by their different dynamics. One of the processes is the relaxation of the excited SL1 state to the ground state of the oxygen-vacancy complex, the other a charge transfer between $^{31}$P donors and SL1 centers forming close pairs, as indicated by electrically detected electron double resonance. For both processes, the recombination dynamics are studied with pulsed EDMR techniques. We demonstrate the feasibility of true zero-field cw and pulsed EDMR for spin 1 systems and use this to measure the lifetimes of the different spin states of SL1 also at vanishing external magnetic field.
\end{abstract}


\maketitle

\section{Introduction\label{secIntro}}

Irradiation of silicon and other semiconductors by high-energy photons or massive particles leads to changes in the electronic and structural properties of these materials which can be attributed to the creation of a large variety of defects. An exposure of semiconductor devices to radiation indeed occurs in different environments such as, e.g., in space applications,\cite{yamaguchi_deep_1999,karazhanov_effect_2000} high-energy physics experiments,\cite{bruzzi_radiation_2001} and nuclear reactors.\cite{leroy_particle_2007} In addition, these radiation-induced defects may be considered as prototype examples of defects created during the fabrication of semiconductor devices, e.g., by ion implantation.\cite{troxell_ion-implantation_1983} To give a concrete example, the oxygen-vacancy center (OV) is among the dominant defects observed in commonly used Czochralski-grown (CZ) oxygen-rich silicon after irradiation with high-energy electrons or $\gamma$-rays.\cite{watkins_spin_1959, hill_electron_1959} Under illumination, the OV center is excited into a metastable triplet state (called SL1) which has gained considerable interest, e.g., for the hyperpolarization of $^{29}$Si nuclear spins \cite{vlasenko_excited_1983, itahashi_optical_2013} and, more recently, in the context of quantum information processing where such excited triplet states have been investigated as mediators coupling nuclear spins.\cite{stoneham_optically_2003, schaffry_entangling_2010} In particular, the SL1 center has been used to demonstrate a coherent state transfer between electron and nuclear spins in silicon.\cite{akhtar_coherent_2012}

The SL1 center has been extensively studied by continuous-wave (cw) and pulsed electron paramagnetic resonance (EPR) methods.\cite{brower_electron_1971, akhtar_rabi_2012} Read-out of the SL1 spin state is also possible by electric means using a spin-dependent recombination process involving the SL1 center.
Two experimental techniques are used for this, spin-dependent conductivity (SDC),\cite{vlasenko_electron_1995} where the conductivity changes are monitored via microwave (mw) reflectivity, and electrically detected magnetic resonance (EDMR), \cite{vlasenko_electron_1995, akhtar_electrically_2009} where the conductance is measured via a dc current flowing through the sample. Both techniques allow to detect much smaller numbers of defects compared to EPR due to their higher sensitivity. While the mechanism leading to spin-dependent recombination at SL1 centers in undoped $\gamma$-irradiated silicon is well understood, \cite{vlasenko_spin-dependent_1986} the situation in irradiated phosphorus-doped silicon has not been studied in detail so far. 
A strong increase of the $^{31}$P EDMR signal intensity has been observed after irradiation with 2 MeV electrons, which was attributed to the formation of $^{31}$P--OV spin pairs.\cite{stich_electrical_1995, akhtar_electrically_2009} However, the presence of such spin pairs has only been inferred indirectly from the simultaneous increase of the SL1 and $^{31}$P EDMR signal amplitude after electron irradiation. In addition, details of this recombination process such as the identity of the involved intermediate states and the time constants of the recombination steps are still unknown.

The spin-dependent recombination processes via $^{31}$P--defect spin pairs are of considerable interest for the read-out of the spin state of donors in silicon.\cite{kane_silicon-based_1998, stegner_electrical_2006, dreher_nuclear_2012}  
So far, such studies have been mostly limited to $^{31}$P--P$_{\mathrm{b}0}$ spin pairs, where P$_{\mathrm{b}0}$  is a dangling bond-type defect at the Si/SiO$_2$ interface.\cite{stesmans_electron_1998}
Spin pairs involving the SL1 are expected to be less susceptible to spurious effects caused by the presence of the Si/SiO$_2$ interface like strain \cite{huebl_phosphorus_2006} and decoherence.\cite{Schenkel2006, de_sousa_dangling-bond_2007} In addition, in contrast to the dangling bond interface defects, the SL1 has a spin of $S=1$ and therefore can be regarded as a model system for the spin-dependent recombination involving other $S=1$ defects like, e.g., the NV--center in diamond,\cite{jelezko_single_2006} potentially aiding the development of electrical rather than optical read-out schemes for this center as well.

Here, we will present a systematic study of the spin-dependent effects on the photoconductivity in $\gamma$-irradiated phosphorus-doped CZ silicon by cw and pulsed EDMR. We will show that two different recombination processes are observed in EDMR, which can be separated due to their different time constants, and will identify them as the spin-dependent recombination via the SL1 only and via a $^{31}\mathrm{P}$--SL1 pair process. A resonant enhancement of the conductivity is observed in connection with the pair process also when switching off the illumination during detection, which can be exploited to improve the signal-to-noise ratio of the $^{31}$P and SL1 spin state read-out.

The paper is organized as follows: In Sect.~II and III we summarize the basic properties of the SL1 center and the experimental methods used for the pulsed EDMR measurements. We discuss the cw (Sect.~IV) and pulsed (Sect.~V) EDMR measurements of a $\gamma$-irradiated phosphorus-doped Si sample.
Detailed discussions of the SL1-only and the $^{31}$P--SL1 pair recombination follow in Sect.~VI and VII, respectively. The presence of these spin pairs and details of the recombination process are in particular studied using electrically detected electron electron double resonance (EDELDOR).\cite{hoehne_spin-dependent_2010} Section VIII is devoted to the determination of the different time constants governing the recombination.
Finally we demonstrate cw and pulsed zero-field EDMR experiments and determine the lifetimes of the SL1 triplet states also in the absence of an external magnetic field (Sect.~IX).

\section{The SL1 Center\label{secSL1}}

\begin{figure}%
\includegraphics[width=\columnwidth]{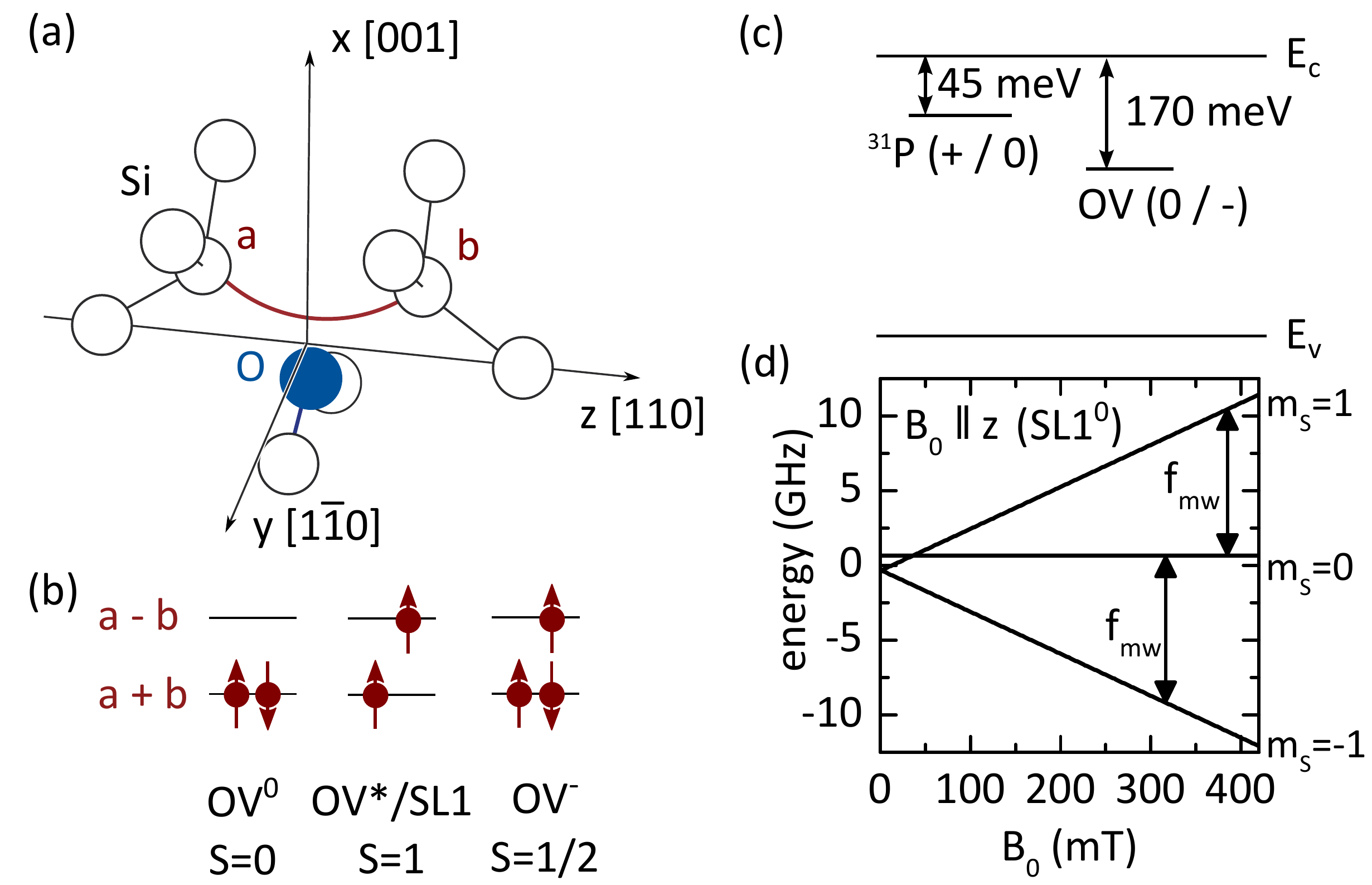}%
\caption{(a) Ball-and-stick model of the OV complex. (b) Schematic representation of the OV complex in its ground state, (excited) triplet state, and negatively charged state. (c) Position of the SL1 0/- charge transfer level. The $^{31}$P +/0 level is shown for comparison.\cite{pearson_electrical_1949} (d) Energy level diagram for the SL1 center (spin $S=1$) as a function of the magnetic field $B_0$.\cite{brower_electron_1971}}%
\label{figSL1}%
\end{figure}

The OV center in silicon is created when a vacancy produced by high-energy irradiation is trapped by an oxygen impurity.\cite{corbett_defects_1961} Its structure is shown in the ball-and-stick model in Fig.~\ref{figSL1}(a). To good approximation, it can be described by a Si--O--Si group (blue lines) which saturates two of the four silicon bonds of the vacancy, while the other two form a weak Si--Si bond (red line) with bonding and antibonding orbitals $a+b$ and $a-b$, respectively.\cite{ortega-blake_electronic_1989, van_oosten_metastable_1994} In the neutral charge state, these are occupied by two electrons which in the ground state OV$^0$ form a spin singlet [Fig.~\ref{figSL1}(b)]. The negative charge state OV$^-$, also referred to as A center, has one half-filled orbital with a spin of $S=1/2$ and has been identified by conventional EPR \cite{watkins_defects_1961} and infrared absorption \cite{corbett_defects_1961} studies. The charge transfer level 0/- is located $170$ meV below the conduction band edge $E_{\mathrm{c}}$  [Fig.~\ref{figSL1}(c)].\cite{wertheim_energy_1957, hill_electron_1959} Also the positive charge state OV$^+$ (charge transfer level +/0 at $E_{\mathrm{v}}+260$ meV, where $E_{\mathrm{v}}$ is the valence band edge)\cite{wertheim_electron-bombardment_1958} has been observed in EPR. \cite{almeleh_electron_1966, brosious_epr_1976}
Because of the rather deep position of these charge transfer levels within the silicon bandgap, the OV acts as an efficient recombination center for Shockley-Read-Hall recombination. \cite{shockley_statistics_1952, baicker_recombination_1963} 

Under above-bandgap illumination the $S=1$ center labeled SL1 has been detected by EPR and identified as the metastable excited triplet state OV$^{\ast}$ of the neutral defect.\cite{brower_electron_1971} The corresponding spin Hamiltonian $\mathcal{H}_s$ for an external magnetic field $\vec{B}_0$ is given by
\begin{align}
	\mathcal{H}_s= \mu_B \vec{B}_0g\vec{S} + \tfrac{1}{2} h\vec{S}D\vec{S}
\end{align}
where $\mu_{\mathrm{B}}$ is the Bohr magneton, $h$ is the Planck constant, $\vec{S}$ is the spin vector for $S=1$, and $g$, describing the electronic Zeeman interaction, and the zero-field splitting tensor $D$ are three-dimensional tensors.
In the defect's native coordinate system, defined as shown in Fig.~\ref{figSL1}(a), $g$ and $D$ are diagonal with $g_{xx}=2.0075$, $g_{yy}=2.0057$, $g_{zz}=2.0102$ and $D_{xx}=700.5$ MHz, $D_{yy}=614.4$ MHz, $D_{zz}=-1314.9$ MHz.\cite{brower_electron_1971}  In Fig.~\ref{figSL1}(d), the eigenenergies of $\mathcal{H}_s$ are shown as a function of a magnetic field $B_0$ in $z$-direction, the two allowed EPR transitions for a mw frequency $f_{\mathrm{mw}}=9.7$ GHz are indicated by arrows. For a magnetic field of $B_0\approx 350$ mT, the first term of $\mathcal{H}_s$ dominates and the triplet states $T_+$, $T_0$ and $T_-$ are in good approximation equal to the eigenstates of the spin operator $S_z$ with eigenvalues $m_S=+1$, $m_S=0$ and $m_S=-1$, respectively. 

\section{Sample and Methods\label{secSample}}

The sample used in this work is a CZ-grown phosphorus-doped silicon sample (size $10\times 4\times 0.7$ mm$^3$) with $\mathrm{[P]}\approx10^{15}$~cm$^{-3}$ as determined by conventional EPR.
For electrical measurements, ohmic contact were made by arsenic implantation in two square-like contact areas with a distance of about $7$ mm and post-implantation annealing. The sample was irradiated with $\gamma$-rays from a $^{60}$Co source with a dose of 3$\cdot 10^{15}$~cm$^{-2}$ resulting in $\mathrm{[SL1]}\approx 2\cdot 10^{11}$~cm$^{-3}$ under illumination, again as determined by EPR, much lower than the concentration of $^{31}$P donors. Then, a Pd-Au layer (with thicknesses of 3~nm and 100~nm, resp.) was evaporated on the implanted areas and contacted via silver paste and Au bonding wires. This contact geometry was favored over interdigit structures \cite{stegner_electrical_2006} to probe the volume rather than the surface of the sample.

All experiments were performed at a temperature of 5\,K in a dielectric mw resonator for pulsed electron nuclear double resonance (ENDOR). The samples were illuminated with the light of a red LED (wavelength of $\sim635$~nm) and biased with 3~V for the cw measurements and 1~V for the pulsed measurements resulting in photocurrents of $I=37.5~\mu$A and $I=16~\mu$A, respectively. The bias points were chosen such that the best signal-to-noise ratios were obtained.

\section{Continuous wave EDMR\label{secCw}}

\begin{figure}
\includegraphics[width=\columnwidth]{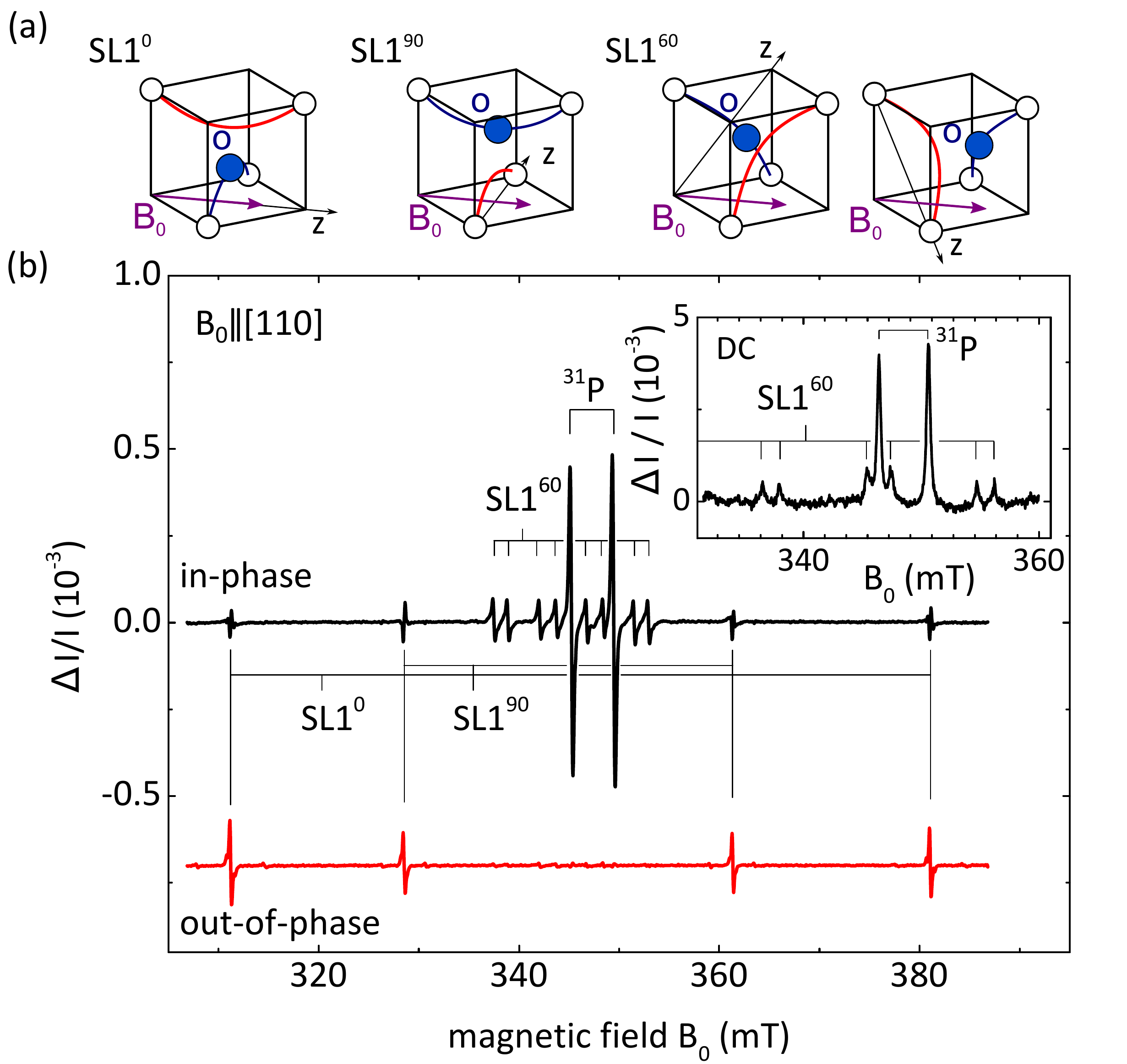}%
\caption{(a) Illustrations of four different orientations of the SL1 complex with respect to the magnetic field. (b) Continuous-wave EDMR spectrum of a phosphorus-doped $\gamma$-irradiated silicon sample. The two traces show the in-phase (black) and out-of-phase (red) signals detected by the lock-in amplifier. The labels SL1$^0$, SL1$^{60}$ and SL1$^{90}$ denote centers oriented along $\langle 110\rangle$ axes enclosing angles of $0^{\circ}$, $60^{\circ}$ and $90^{\circ}$ with the magnetic field, resp., as shown in (a). The inset shows the central part of the EDMR spectrum recorded in dc mode without modulation and reveals the $^{31}\mathrm{P}$ and SL1$^{60}$ resonances as a resonant increase of the photocurrent.}%
\label{cwSpec1}%
\end{figure}

In a first step, we characterized the sample using cw EDMR. The sample was irradiated with a mw frequency $f_\mathrm{mw}=9.7$ GHz at a power of $40$~mW with the mw resonator tuned to a high quality factor. The resulting relative change of the photocurrent $\Delta I/I$ is shown in Fig.~\ref{cwSpec1}(b) for the in-phase (black) and out-of-phase (red) signals of the lock-in amplifier for a peak-to-peak magnetic field modulation of $\Delta B_0=0.2$~mT at a frequency $f_{\mathrm{mod}}= 500$ Hz. The phase has been adjusted such that the two strongest peaks ($B_0=345.2$ mT and $B_0=349.4$ mT) are observed in the in-phase signal only. They are associated with the two hyperfine-split lines of the phosphorus donor electron with a characteristic hyperfine splitting of $4.2$ mT. \cite{feher_electron_1959} The spectral positions of the 12 smaller lines in the upper trace correspond well to the spectral positions expected for the SL1 center for $B_0$ parallel to the $[110]$ crystal direction with an additional tilt of $\approx2^{\circ}$ about the [100] direction. \cite{brower_electron_1971} The SL1 peaks at different spectral positions can be associated with SL1 centers with different orientations with respect to $B_0$. The labels SL1$^{0}$, SL1$^{60}$, and SL1$^{90}$ denote SL1-centers oriented along $\langle 110 \rangle$ directions enclosing angles of 0$^\circ$, 60$^\circ$, and 90$^\circ$ with the magnetic field [Fig.~\ref{cwSpec1}(a)], respectively. With a value of $\Delta I/I$ of about $0.5\cdot 10^{-3}$, the signal intensity in this experiment is significantly higher than that found for corresponding experiments at Si/SiO$_2$ interfaces.\cite{stegner_electrical_2006, huebl_phosphorus_2006}

In the out-of-phase signal of the lock-in amplifier only the SL1$^{90}$ and SL1$^{0}$ peaks are observed while in the in-phase signal all SL1 peaks appear with similar intensities. In magnetic field-modulated EDMR, the observation of signals with different phases can be related to spin-dependent transport processes with different characteristic time constants.\cite{dersch_recombination_1983} The experimental data in Fig.~\ref{cwSpec1}(b) therefore suggest that two different processes are observed in this sample, one process with a slower time constant which involves the SL1 center only, and another process with a shorter time constant involving $^{31}\mathrm{P}$ donors and SL1 centers. The signal intensity of the former shows a pronounced dependence on the orientation of the SL1 center with respect to the magnetic field. This is visible in Fig.~\ref{cwSpec1}(b) where only the SL1$^{0}$ and SL1$^{90}$ orientations are observed for the out-of-phase signal, while all SL1 and $^{31}$P lines are observed in the in-phase signal.

To determine the sign of the resonant changes of the conductivity, EDMR without magnetic field modulation, where the dc current is recorded directly without lock-in amplification, was performed. As shown in the inset of Fig.~\ref{cwSpec1}(b), the center lines correspond to enhancements in the photocurrent. This observation is surprising since in photo-EDMR experiments, mw-induced transitions typically enhance the recombination rate, hence decreasing the carrier mobility resulting in a decrease of the photocurrent. \cite{lepine_spin-dependent_1972, kaplan_explanation_1978} The origin of the resonant current increase will be addressed in Sect.~VII C. 
Because the sample was reinstalled between experiments, the orientation of the sample differs slightly for the measurements shown in the inset of Fig. \ref{cwSpec1}(b) and those in the main part of that figure, which is particularly noticed in the position of the SL1$^{60}$ peaks. The same holds for the data in Fig.~\ref{pEDMRSpec}(b) and Fig.~\ref{figDarkLight}(a).

\section{Pulsed EDMR\label{secPEDMR}}
To further elucidate the different behaviour of the resonances with respect to modulation and magnetic field orientation, we use pulsed EDMR~\cite{boehme_theory_2003,stegner_electrical_2006} to separate them in the time domain.
To this end, the sample was irradiated with a $90$ ns mw pulse at frequency $f_{\mathrm{mw}}=9.74$ GHz and mw power $P\approx 50$ W with the resonator set to a low quality factor. The photocurrent transient after the pulse was recorded under continuous illumination for different values of the magnetic field $B_0$ in [110] direction.
\begin{figure}%
\includegraphics[width=\columnwidth]{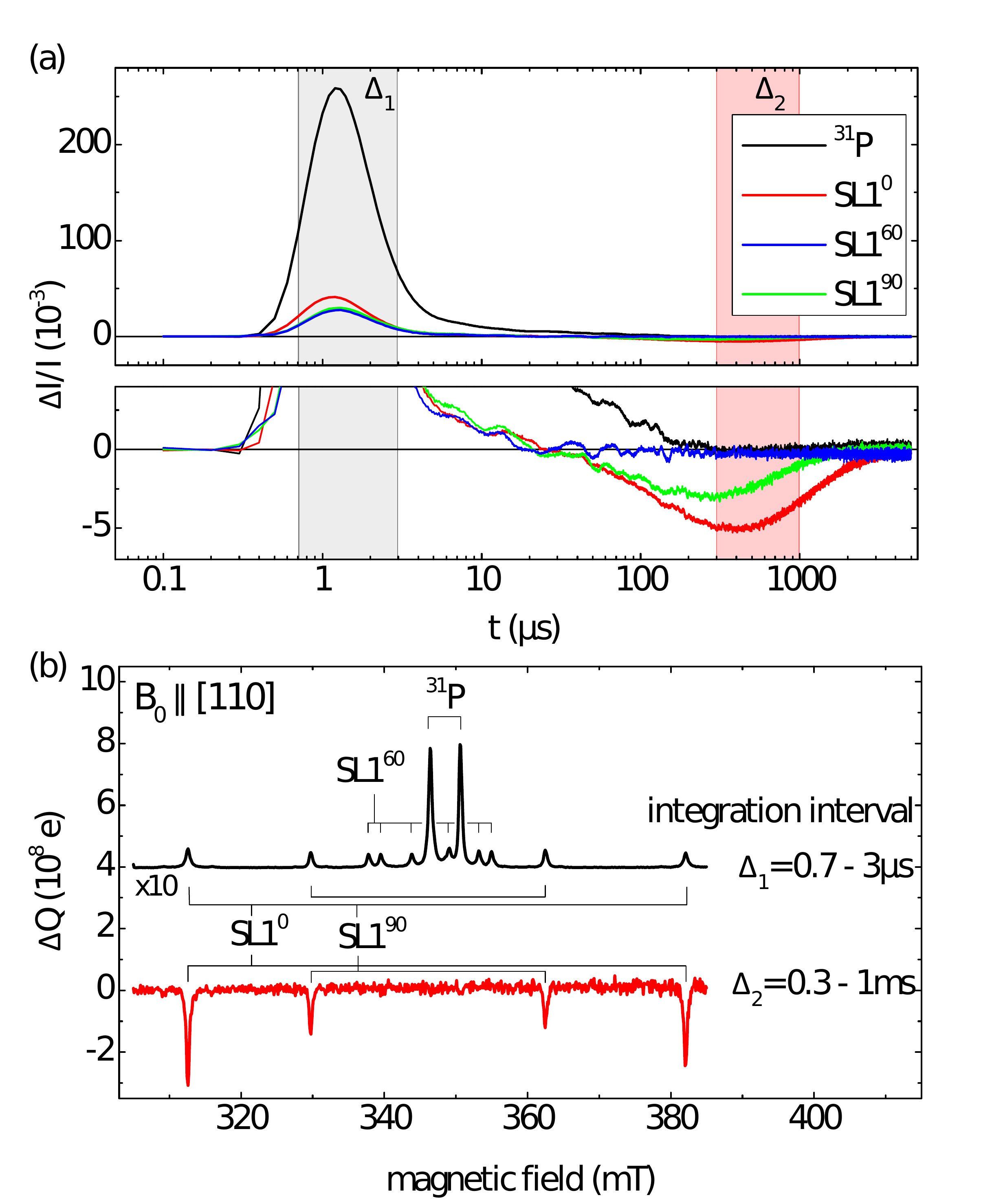}%
\caption{(a) Change in photocurrent after a $90$ ns mw pulse for different magnetic fields $B_0$ of $312.6$ mT (SL1$^0$), $329.7$ mT (SL1$^{90}$), $339.5$ mT (SL1$^{60}$) and $346.4$ mT ($^{31}$P) after subtraction of the non-resonant background. (b) Pulsed EDMR spectra for two different integration intervals $\Delta_1$ (black) and $\Delta_2$ (red) as indicated in (a). }%
\label{pEDMRSpec}%
\end{figure}

The resulting transients are shown in Fig.~\ref{pEDMRSpec}(a) where the normalized resonant change in photocurrent $\Delta I/I$ is shown as a function of the time $t$ after the mw pulse for the low-field $^{31}\mathrm{P}$ peak and three additional peaks corresponding to the three different SL1 orientations. For all peaks, the current first increases with a timeconstant of approximately $0.7~\mu$s followed by a decrease for $t\approx 2~\mu$s.  
For SL1$^{0}$ and SL1$^{90}$ (green and red traces), an additional transient with slower time constants of the order of several $100~\mu$s and a negative $\Delta I/I$ is observed.

The two different time scales of the two recombination processes are revealed more clearly by plotting the integrated pulsed EDMR transients $\Delta Q$ as a function of the magnetic field for two different box-car integration intervals $\Delta_1=0.7-3~\mu$s and $\Delta_2=0.3-1$ ms [Fig.~\ref{pEDMRSpec}(b)]. For $\Delta_1$, the spectrum shows a positive signal for all SL1 peaks and the $^{31}\mathrm{P}$ peaks with similar intensities for all SL1 lines.
In contrast, for the integration interval $\Delta_2$, peaks with a negative $\Delta Q$ are observed for SL1$^0$ and SL1$^{90}$, while for $^{31}$P and SL1$^{60}$ no signals are visible. This directly confirms the observations made in the cw EDMR measurements and allows to separate the signals from the two recombination processes in the time domain by choosing the appropriate integration interval.

\section{Spin-dependent recombination via the OV defect only\label{secSDR}}
 The spectra associated with the slower process (out-of-phase signal in cw EDMR, integration interval $\Delta_2$ in pulsed EDMR) shows notable similarities to the SDC spectra of undoped $\gamma$-irradiated CZ-silicon. In the following we will identify it as the spin-dependent recombination process observed in SDC and leading to the polarization observed in EPR.\cite{vlasenko_excited_1983,vlasenko_epr_1984, vlasenko_electron_1995}
 
\begin{figure}%
\includegraphics[width=\columnwidth]{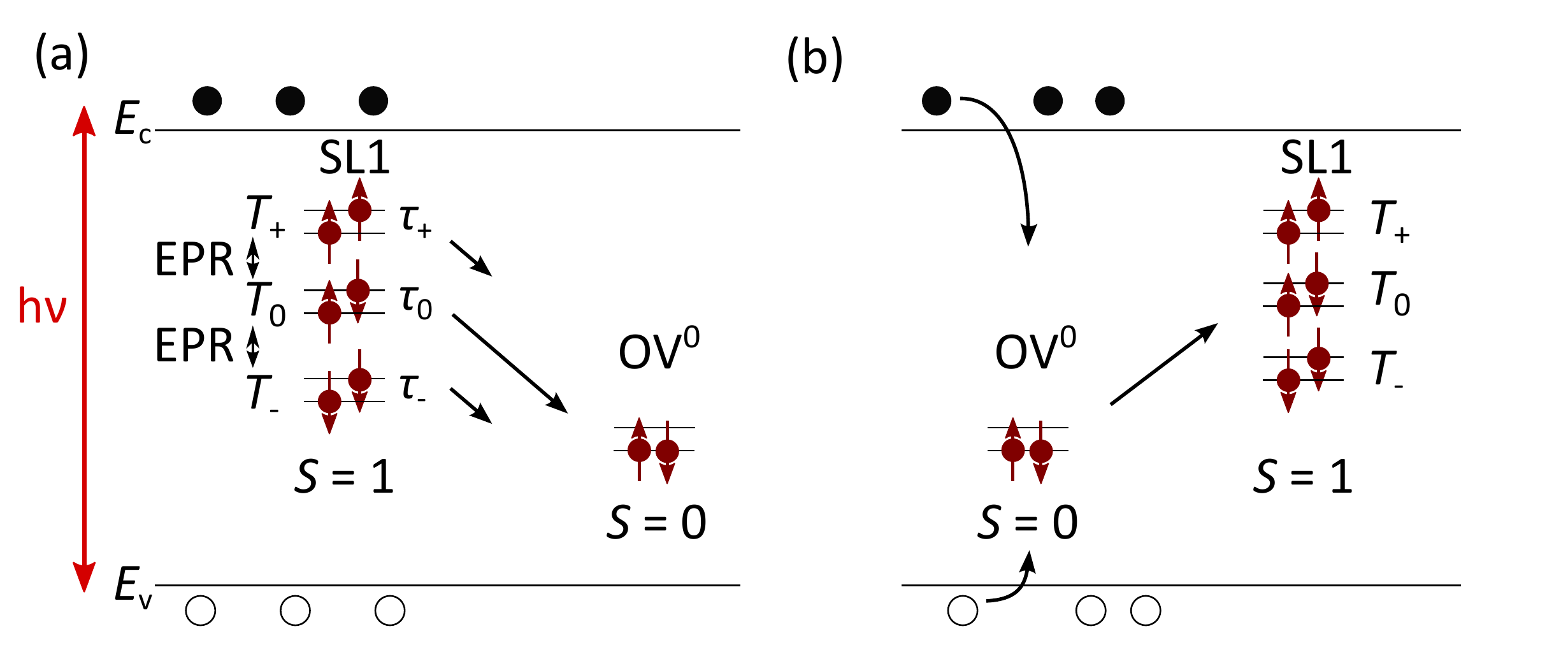}%
\caption{Spin-dependent recombination process via the excited triplet state of the OV complex. (a) Spin-orbit interaction leads to the relaxation of the photoexcited triplet states to the singlet ground state OV$^0$ with lifetimes $\tau$. Inducing EPR transitions between these states leads to more defects in the states with shorter lifetime and therefore promotes the relaxation. (b) The OV$^0$ acts as an efficient carrier trap, reducing the number of charge carriers. This change can be observed as a reduction of photocurrent in EDMR. On subsequent capture of an electron and a hole, the defect can be excited to the triplet state SL1, completing the cycle. For clarity, $T_0=1/\sqrt{2}(\mid\uparrow\downarrow\rangle +\mid\downarrow\uparrow\rangle)$ is displayed in one possible projection.}%
\label{figRec_process}%
\end{figure}

In this model, summarized in Fig.~\ref{figRec_process}, the successive capture of an electron $e$ and a hole $h$ leads to the formation of the excited triplet state SL1. The excitation process can be written as
\begin{align}
	\text{OV}  \left\{\begin{array}{cc}
	\stackrel{+e}{\longrightarrow} \text{OV}^- \stackrel{+h}{\longrightarrow}\\
	\stackrel{+h}{\longrightarrow} \text{OV}^+ \stackrel{+e}{\longrightarrow}
	\end{array}\right\} \rightarrow \text{OV}^{\ast} / \text{SL1}\text{ ,}
\end{align}
where OV$^-$ and OV$^+$ denote the negative and positive charge states of the OV complex, respectively.
The excited triplet state is long lived because the transition to the singlet ground state is mediated by the weak spin-orbit coupling. The probability of this transition strongly depends on the spin projection $m_S$ and is in general different for the SL1 triplet states $T_0$ and $T_{\pm}$ [Fig.~\ref{figRec_process}(a)]. Since the formation rate for all three triplet states is assumed to be equal, the states with slower recombination rates have higher equilibrium populations compared to the faster recombining states.

Inducing transitions between the triplet states via EPR enhances the number of centers in the faster recombining states and, therefore, the average life time of the SL1 is decreased. Since the singlet ground state of the OV complex is an efficient recombination center,\cite{baicker_recombination_1963} this leads to a higher recombination rate for charge carriers in the conduction and valence bands [Fig. \ref{figRec_process}(b)] resulting in the decrease of the photoconductivity which is observed in EDMR measurements.

The different recombination rates for the triplet states $T_{\pm}$ and $T_0$ can be estimated from perturbation theory and strongly depend on the orientation of the external magnetic field.\cite{vlasenko_epr_1984} For the SL1$^{60}$ orientation, the recombination rates for the three triplet states are almost equal, so that all states are equally populated and are almost not altered by resonant mw irradiation. We therefore do not expect an EDMR signal, which is in agreement with the observations in Fig.~\ref{cwSpec1}(b) and Fig.~\ref{pEDMRSpec}(b). In contrast, for the SL1$^0$ and SL1$^{90}$, the lifetimes of the $T_\pm$ and the $T_0$ states differ more strongly, \cite{vlasenko_epr_1984} so that a resonant decrease of the photoconductivity is observed.
We will use pulsed EDMR in combination with pulsed illumination \cite{hoehne_time_2013} to determine the characteristic time constants of this recombination process as discussed in detail in Sect.~VIII.

\section{Spin-dependent $^{31}$P--SL1 pair recombination process}
\subsection{EDELDOR experiments}\label{chapELDORELDOR}

Several observations suggest that the signal associated with the fast transient observed in the integration interval $\Delta_1$ involves the formation and recombination of $^{31}$P-SL1 spin pairs. First, the shapes of the faster current transients in the pulsed EDMR measurements are similar for the $^{31}$P peak and the SL1 peaks [cf.~Fig.~\ref{pEDMRSpec}(a)] indicating very similar dynamics. Second, the sum of the integrated signal amplitudes of the two $^{31}$P peaks is almost equal to the sum of all integrated SL1 signal amplitudes. The relevance of $^{31}$P-SL1 spin pairs is further supported by the increase of the EDMR $^{31}$P signal intensity in P-doped silicon after irradiation with high-energy electrons observed in Ref.~\onlinecite{stich_electrical_1995}. 

\begin{figure}%
\includegraphics[width=\columnwidth]{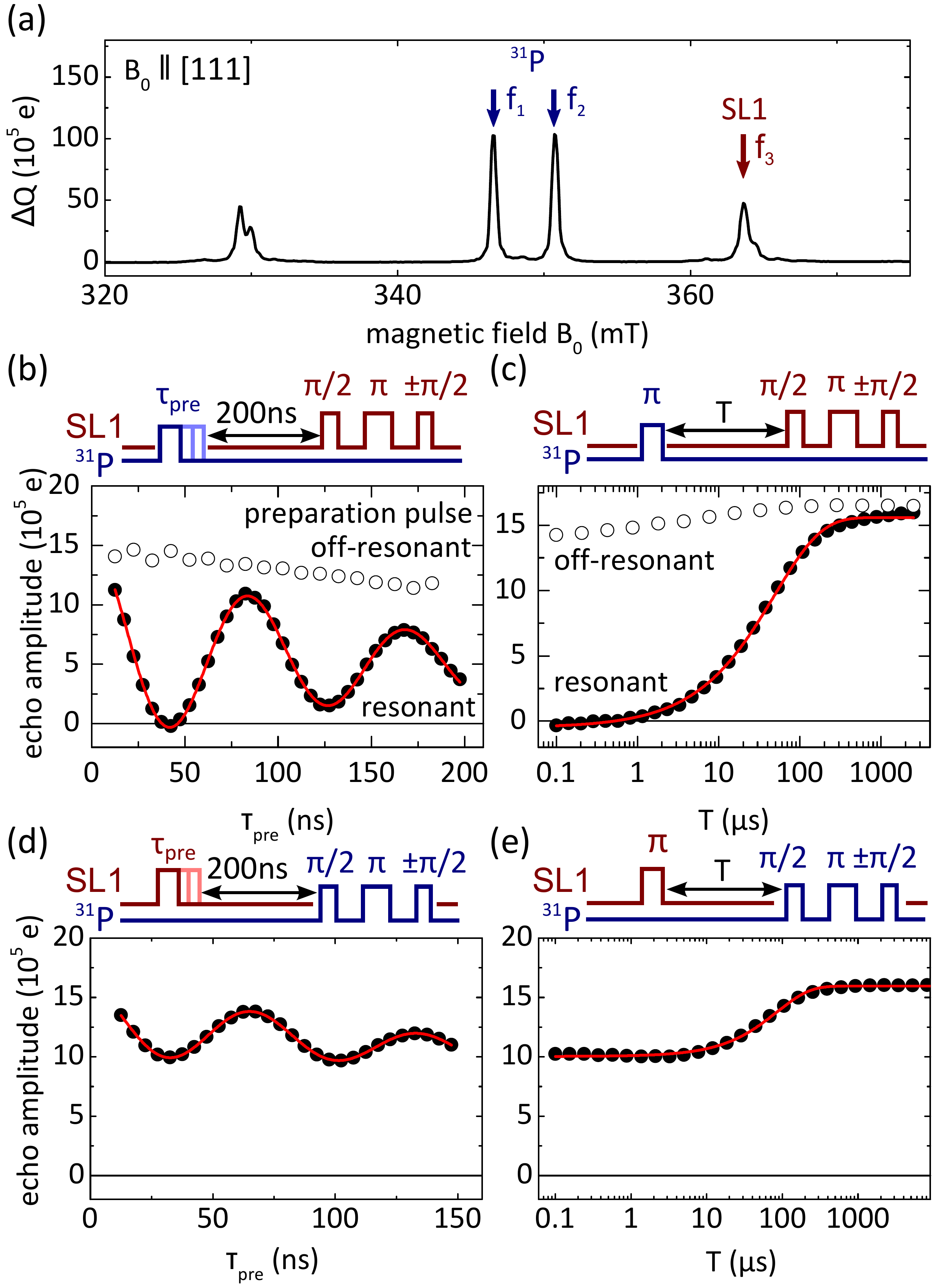}%
\caption{
(a) Pulsed EDMR spectrum showing the three transitions used in the EDELDOR experiments.
(b) SL1 echo amplitude recorded as a function of the length of the $^{31}$P preparation pulse (at $f_1$ and $f_2$) and (c) as a function of the time $T$ between the $^{31}$P inversion pulse and the spin echo.
(d) $^{31}$P echo amplitude (at $f_2$) as a function of the length of the SL1 preparation pulse (at $f_3$) and (e) as a function of the time $T$ between the SL1 inversion pulse and the spin echo.
Red lines are fits with a damped sine function in (b) and (d) and with an exponential function in (c) and (e). All EDELDOR measurements were performed under continuous illumination.}
\label{figELDOR1}%
\end{figure}

In the following, we use EDELDOR \cite{hoehne_spin-dependent_2010} to directly demonstrate the presence of $^{31}$P-SL1 spin pairs. In EDELDOR, mw pulses of different frequencies are used to selectively excite transitions for the two partners in a pair recombination process. Starting from a long-lived $^{31}$P-SL1 spin pair in the steady state under illumination, we first alter the spin pair symmetry by a mw preparation pulse of length $\tau_{\mathrm{pre}}$, changing the $^{31}$P spin state [blue trace in the pulse sequences of Fig.~\ref{figELDOR1}(b) and \ref{figELDOR1}(c)]. After a time interval $T$, we monitor the state of the spin pair using a spin echo sequence, including a final projection pulse,\cite{huebl_spin_2008} on the SL1 transition [red trace in the pulse sequence]. If the spin-dependent EDMR signal is determined by the symmetry of the spin pair, we expect a variation of the spin echo amplitude for different preparation pulse lengths $\tau_\mathrm{pre}$. In contrast, if the $^{31}$P and SL1 EDMR signals originate from two independent recombination processes no such change is expected.

We use two mw frequencies $f_1$ and $f_2$ to simultaneously excite both $^{31}$P transitions as indicated by the blue arrows in the corresponding spectrum shown in Fig.~\ref{figELDOR1}(a). The chosen mw power resulted in a $\pi$ pulse length of $\tau_{\pi}=42.5$~ns for the preparation pulse.
A third frequency $f_3$ is chosen to match an SL1 resonance [red arrow in Fig.~\ref{figELDOR1}(a)]. The orientation of the sample has to be chosen such that one SL1 resonance is spectrally close enough to the $^{31}$P lines to excite all three lines within the bandwidth of the mw resonator, but spectral overlap is avoided. This was achieved for an angle of $\sim 20^{\circ}$ of the $[110]$ axis with the external field. For the detection spin echo at $f_3$, a two-step phase cycle sequence is employed to implement a lock-in detection scheme where the phase of the last $\pi$/2 pulse is switched by 180$^\circ$ as indicated by the $\pm\pi/2$ in Fig.~\ref{figELDOR1}(b)-\ref{figELDOR1}(e). For details see Ref.~\onlinecite{hoehne_lock-detection_2012}. In the present work, the sign of the subtraction of the signal for the two cycles is chosen differently from Ref.~\onlinecite{hoehne_lock-detection_2012} so that the sign of the echo amplitude represents the sign of the change in photocurrent in the pulsed EDMR spectrum.

In Figure~\ref{figELDOR1}(b), the detection echo amplitude on the SL1 transition is shown as a function of the preparation pulse length $\tau_\mathrm{pre}$ for $T=200$ ns (full circles). A pronounced oscillation with a period of $\approx$85~ns is visible with a complete quenching of the echo amplitude for, e.g., $\tau_\mathrm{pre}$=42.5~ns corresponding to a preparation $\pi$ pulse. In contrast, if the frequencies of the preparation pulse are detuned from the $^{31}$P transitions no oscillation is observed (open symbols). The dynamics of the quenching can be studied by varying the time $T$ between inversion pulse and detection echo while choosing $\tau_\mathrm{pre}=\tau_\pi$, as shown in the pulse sequence in Fig.~\ref{figELDOR1}(c). The quenching of the echo signal is observed even for the shortest interval $T=100$ ns investigated which is more than one order of magnitude shorter than the antiparallel recombination time $\tau_{\mathrm{ap}}\approx 4~\mu$s (cf.~Sect.~VIII). From this observation it is clear that a change in the spin symmetry of the pair leads to a change in the signal for both partners which is observed even before a recombination has taken place.
For larger time intervals $T$, the echo amplitude recovers with a characteristic time constant of $49\pm 10~\mu$s [Fig.~\ref{figELDOR1}(c)], which is interpreted as the pair generation rate governing the recreation of pairs after the recombination. The quenching of the echo amplitude directly after the preparation pulse suggests that a direct spin-dependent electron transition between the $^{31}$P donor a the SL1 is observed without involving an intermediate state such as, e.g., an OV$^+$. 

To further support our assignment to a  $^{31}$P--SL1 spin pair, we repeated the experiments described above, but reversed the roles of the $^{31}$P and the SL1 applying the preparation pulse on the SL1 transition and the detection echo on one of the $^{31}$P transitions. The sample was oriented with a $\langle 111 \rangle$ crystal axis in direction of the magnetic field $B_0$, so that a maximum spectral overlap of the SL1 lines was achieved [Fig.~\ref{figELDOR1}(a)]. The mw power used resulted in a $\pi$ pulse length of $32.5$ ns for the SL1. Again, a pronounced oscillation of the spin echo amplitude as a function of $\tau_\mathrm{pre}$ is observed [Fig.~\ref{figELDOR1}(d)] confirming the spin pair hypothesis. In this case, the echo amplitude is not completely quenched even for a $\pi$ preparation pulse, since this pulse addresses only one of the  two SL1 lines. The bandwidth of the preparation pulse ($\sim 37$ MHz) is larger than the SL1 line width but not large enough to excite the two $^{29}$Si satellite peaks, reducing the maximum possible decrease to $\sim 45$\%. 
The echo amplitude recovers with a characteristic time constant of 85$\pm10$~$\mu$s [Fig.~\ref{figELDOR1}(e)]. The slower pair generation rate observed in this case is caused by the lower illumination intensity in this orientation.

\subsection{Model of the spin-dependent pair process}\label{chapELDORModel}
\begin{figure}%
\includegraphics[width=\columnwidth]{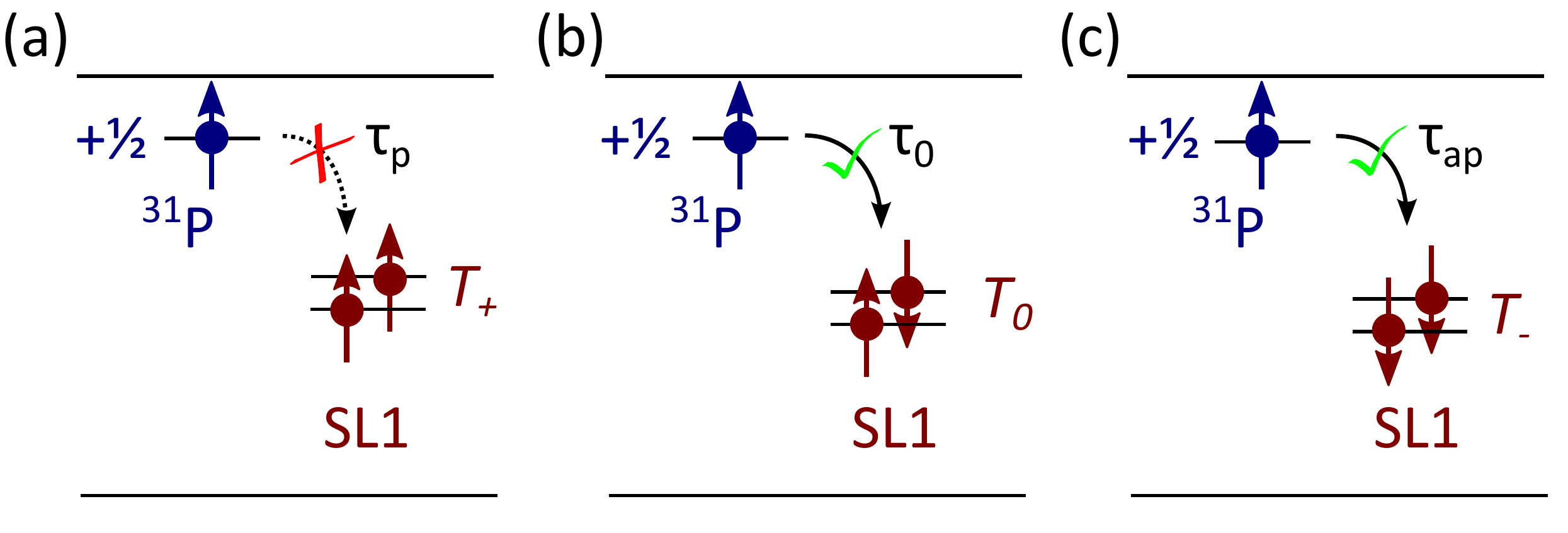}%
\caption{Possible pair configurations for a $^{31}\mathrm{P}$--SL1 pair. (a) For the parallel configuration $\Psi_\mathrm{p}$, the three electrons are in the same spin state and the overall spin wave function is symmetric. A charge transfer would lead to the formation of a symmetric orbital wave function and is therefore forbidden by the Pauli principle. (b), (c) For the mixed and antiparallel configurations $\Psi_0$ and $\Psi_\mathrm{ap}$, resp., the transition is allowed because an antisymmetric spin wave function can be formed. As in Fig. \ref{figRec_process}, just one representation of $T_0$ is shown.}%
\label{figELDORstates}%
\end{figure}
Based on these findings, we interpret the fast spin-dependent recombination process in terms of a donor-acceptor or spin pair model. \cite{stich_electrical_1995,cox_exchange_1978, kaplan_explanation_1978}  The 0/- charge transfer level of the OV is located $170$ meV below the conduction band \cite{watkins_spin_1959} which is below the phosphorus +/0 level ($45$ meV below the conduction band)\cite{pearson_electrical_1949}, so that a charge transfer from the neutral $^{31}\mathrm{P}$ to the neutral OV is possible and has been observed in ODMR experiments.\cite{chen_direct_1991}
In the spin-pair model, the $^{31}$P and the SL1 form a weakly coupled spin pair when the distance between them is small enough to allow a direct transition of the $^{31}$P electron to the SL1, but large enough so that the coupling of the two spins is small compared to the Zeeman energy.

For the $^{31}$P-SL1 spin pair, we consider three possible spin symmetries: a parallel configuration [Fig.~\ref{figELDORstates}(a)]
\begin{align*}
	\mid \Psi_{\mathrm{p}}\rangle = \mid\uparrow, T_+ \rangle\text{, }	 \mid \downarrow, T_- \rangle\text{,}
\end{align*}
a mixed configuration [Fig.~\ref{figELDORstates}(b)]
\begin{align*}
	\mid \Psi_{0}\rangle = \mid\uparrow, T_0 \rangle\text{, }	 \mid \downarrow, T_0 \rangle\text{,}
\end{align*}
and an antiparallel configuration [Fig.~\ref{figELDORstates}(c)] 
\begin{align*}
	\mid \Psi_{\mathrm{ap}}\rangle = \mid\uparrow, T_- \rangle\text{, }	 \mid \downarrow, T_+ \rangle\text{,}
\end{align*}
where the arrows denote the $^{31}$P spin states and $T_j$ ($j\in\{+,0,-\}$) the SL1 spin states.

\begin{figure*}%
\includegraphics[width=\textwidth]{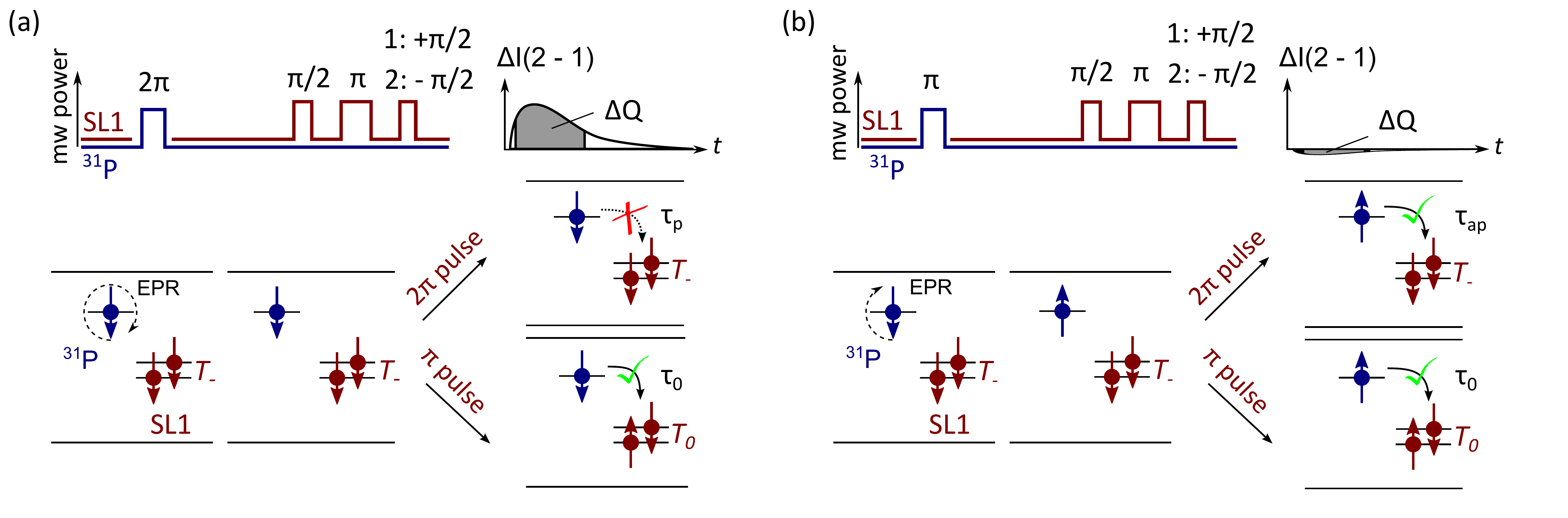}%
\caption{(a) Effect of a spin echo sequence on a spin pair consisting of a $^{31}\mathrm{P}$ electron spin and an SL1 center. (b) Effect of an EDELDOR experiment with an inversion pulse on the $^{31}\mathrm{P}$ resonance and a subsequent spin echo on the SL1 resonance.}%
\label{figELDORpair2}%
\end{figure*}

We further assume that the rates of the electronic transition from the donor to the SL1 center strongly depend on the relative spin orientation. The lifetime $\tau_\mathrm{p}$ for parallel spin pairs is expected to be much larger than the lifetime $\tau_\mathrm{ap}$ for antiparallel spin pairs, which is a consequence of the Pauli principle and will be confirmed by measurements of these rates described in Sect.~VIII. For the mixed configuration the corresponding lifetime is given by 
\begin{align}
	\tau_0=\left(\frac{1}{2\tau_{\mathrm{p}}}+\frac{1}{2\tau_{\mathrm{ap}}}\right)^{-1}\approx 2\tau_{\mathrm{ap}}\text{ ,}
\end{align}
if $\tau_\mathrm{ap} \ll \tau_\mathrm{p}$.
The spin-dependent charge transfer process can be summarized by
\begin{align}
	^{31}\text{P}+\text{SL1} \left\{\begin{array}{cc}
		\mid\Psi_{\mathrm{ap}}\rangle & \stackrel{\tau_{\mathrm{ap}}}{\longrightarrow}\\
		\mid\Psi_{\mathrm{p}}\rangle & \stackrel{\tau_{\mathrm{p}}}{\longrightarrow}\\
		\mid\Psi_{0}\rangle & \stackrel{\tau_{0}}{\longrightarrow}
	\end{array}\right\} {}^{31}\text{P}^+ + \text{OV}^-\text{ .}
\end{align}
We will now discuss the results of the EDELDOR measurements in terms of this model.
We start by considering the case of a 2$\pi$ preparation pulse (or equivalently no preparation pulse) sketched in Fig.~\ref{figELDORpair2}(a). Under illumination the spin pairs will mostly be in the long-lived parallel configuration shown here exemplarily as the $\left|\downarrow , T_-\right\rangle$ state. A 2$\pi$ preparation pulse does not alter the spin pair configuration so that the spin pair is still in the parallel state before the detection spin echo. For the two-step phase cycle the echo forms an effective 2$\pi$ pulse for a final +$\pi$/2 pulse in step 1 and an effective $\pi$ pulse for the -$\pi$/2 pulse in step 2. After an effective 2$\pi$ pulse on the SL1 transition the spin pair is still in the long-lived parallel configuration, while for an effective $\pi$ pulse the spin pair is transferred to the short-lived mixed configuration reducing the spin pair lifetime in this case. Subtracting the box-car integrated current transients for step 1 from 2 yields the echo amplitude $\Delta Q$ observed, e.g., for $\tau_\mathrm{pre}$=85~ns in Fig.~\ref{figELDOR1}(b).

In contrast, for a $\pi$ preparation pulse on the $^{31}$P transition the spin pair is in the antiparallel configuration before the detection spin echo.    
In this case both, the spin echo forming an effective 2$\pi$ pulse (step 1) as well as the echo forming a $\pi$ pulse (step 2) result in short-lived configurations $\Psi_\mathrm{ap}$ and $\Psi_\mathrm{0}$, respectively. Since the difference between the corresponding lifetimes $\tau_\mathrm{ap}$ and $\tau_\mathrm{0}$ is small compared to the difference to $\tau_{\mathrm{p}}$, the resulting current transients for both steps cancel each other when subtracted, so that the echo amplitude $\Delta Q$ is almost zero as observed in the experiment for $\tau_{\mathrm{pre}}$=42.5~ns in Fig.~\ref{figELDOR1}(b). This result differs from what is observed for a spin pair involving two $S=1/2$ spins, where a complete inversion of the echo signal is observed.\cite{hoehne_spin-dependent_2010}

Based on the spin pair model, we expect that the recombination rates only depend on the symmetry of the spin pair and not on the orientation of the SL1 center, so that similar signals are expected for the SL1$^0$, the SL1$^{60}$ and the SL1$^{90}$. This is indeed observed in the cw EDMR and pulsed EDMR experiments, where the signal amplitudes and for the latter also the time constants of the current transients are similar for all SL1 peaks (cf.~Fig.~\ref{pEDMRSpec}). 

\subsection{Enhancement in photocurrent and dark readout mechanism}\label{secEnhancement}

\begin{figure}%
\includegraphics[width=\columnwidth]{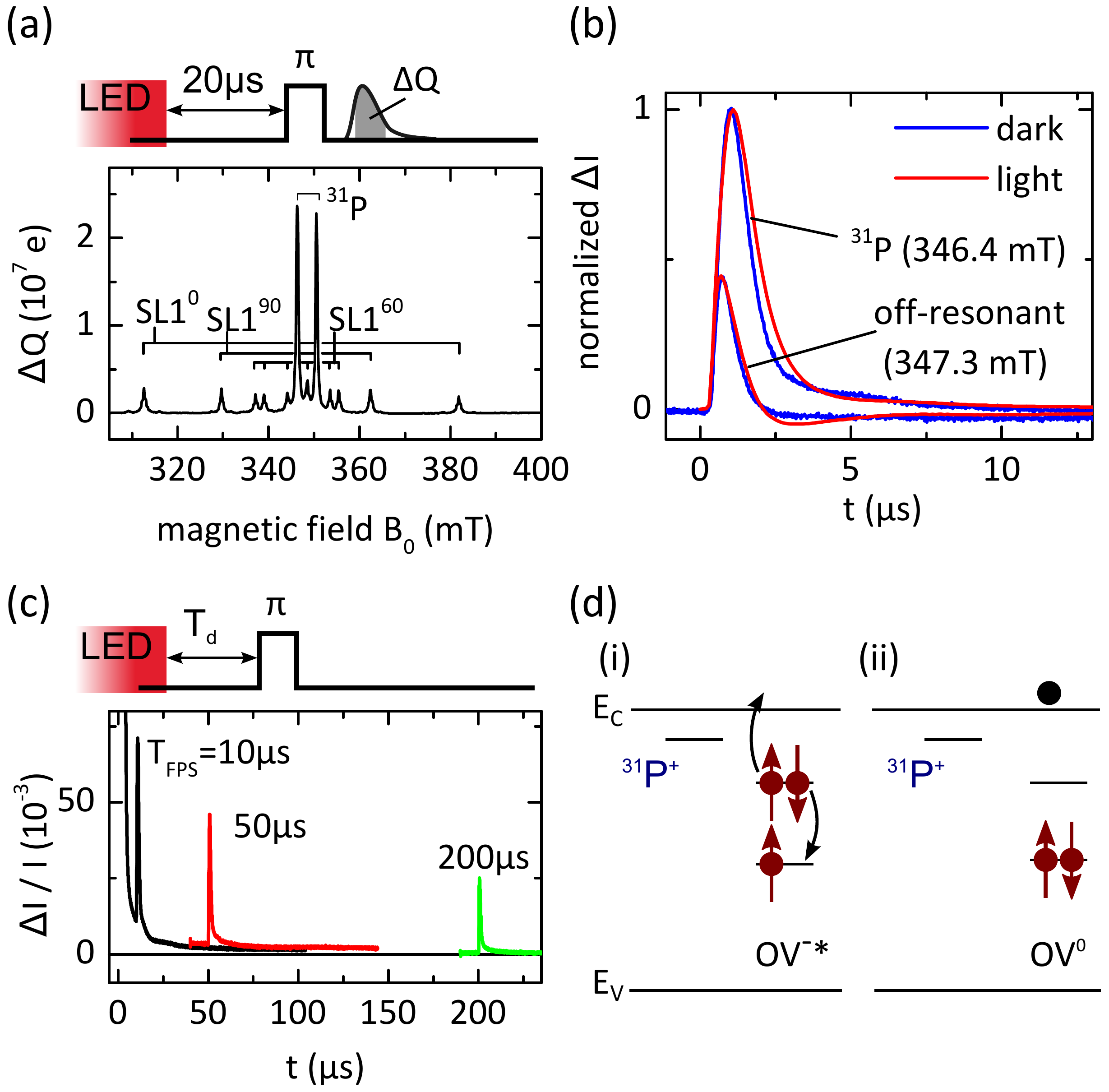}%
\caption{(a) Pulsed EDMR spectrum recorded by application of a mw pulse $20~\mu$s after switching off the illumination and integration of the current transient observed in the dark.
(b) Comparison of resonant and off-resonant dark current transients with current transients recorded under continuous illumination. (c) Transient change in current after a resonant mw pulse ($B_0=364.4$ mT) for different delay times $T_{\mathrm{d}}$. (d) Schematic representation of a possible Auger process. (i) After the charge transfer an excited negative state OV$^{-\ast}$ is formed. (ii) On relaxation to the OV$^0$ ground state, an electron is emitted to the conduction band.}%
\label{figDarkLight}%
\end{figure}

The experiments described so far do not allow to access the last step of the recombination
process, namely the formation of new $^{31}$P--SL1 spin pairs after the spin-dependent charge transfer process creating $^{31}$P$^+$--OV$^-$ pairs and the cause of the observed increase in photocurrent.
After the charge transfer, the phosphorus donor is left in its ionized state and will eventually capture an electron from the conduction band, while the OV$^-$ could either emit an electron into the
conduction band or capture a hole from the valence band. 
Evidence for the emission of an electron is obtained by measuring the current response to a resonant mw pulse applied at a time $T_\mathrm{d}$=20~$\mu$s after switching off the illumination, much longer than the fall time of the illumination pulse of $\approx 3~\mu$s and the carrier lifetime ($<1~\mu$s) [Fig.~\ref{figDarkLight}(a)].
At the time of the mw pulse, nearly all free carriers have recombined and the photocurrent is almost zero. 
Nevertheless, a current transient is observed after the resonant mw pulse, which is much larger compared to the current transient for a non-resonant mw pulse. As shown by the integrated spectrum shown in Fig.~\ref{figDarkLight}(a), this is true for all lines of the $^{31}$P and SL1 spectra demonstrating that the dark current transient is related to the $^{31}$P--SL1 spin-dependent charge transfer process. This is further confirmed by comparing the shape of the resonant and off-resonant current transients in the dark and under illumination, which are nearly identical as shown in Fig.~\ref{figDarkLight}(b) suggesting that both transients originate from the same spin-dependent process. The amplitude of the observed transients is reduced for longer waiting times $T_{\mathrm{d}}$ [Fig.~\ref{figDarkLight}(c)]. More detailed measurements show that the resonant part decays with a timeconstants that we interpret as the parallel lifetime of the spin pair in Sect.~VIII.

From these observations we can exclude the quenching of a competing recombination process as a possible explanation for the current increase as well as a change in the charge carrier mobility. Spin-dependent hopping processes can cause a resonant enhancement of conductivity \cite{Morigaki_spin-dependent_1974, schnegg_pulsed_2012} but seem unlikely because of the comparatively low concentration of donors.

One possible mechanism leading to the observed behaviour would be the excitation of charge carriers in an Auger-type process. If the OV$^-$ is in an excited state after the spin-dependent transition, the energy provided by the relaxation into its ground state could be transferred to one of the electrons in the OV$^-$, thereby exciting it to the conduction band. This is show schematically in Fig.~\ref{figDarkLight}(d). This relaxation process is not forbidden by spin selection rules, therefore possibly making it sufficiently fast to explain the immediate current increase after the mw pulse observed in the experiment. For such an excitation to take place, the energy gained on the relaxation of one electron to the binding orbital $E_{\mathrm{OV}^{\ast}}-E_{\mathrm{OV}^0}$ has to be larger than the ionization energy $E_\mathrm{c}-E_{\mathrm{OV}^{-\ast}}$ of the excited negative state. On the other hand, the OV$^{\ast}$ 0/- level has to be located below the $^{31}$P +/0 level to allow a charge transfer. The SL1 excitation energy has been estimated to be $E_{\mathrm{OV}^{\ast}}-E_{\mathrm{OV}^0}\approx 160$ meV, \cite{ortega-blake_electronic_1989, van_oosten_metastable_1994} resulting in the condition $45 \mathrm{ meV} < E_\mathrm{c}-E_{\mathrm{OV}^{-\ast}} < 160\mathrm{ meV}$ for the assumed OV$^{-\ast}$ 0/- level. However, further experiments will be necessary to confirm this explanation. For example ENDOR measurements of the $^{29}$Si hyperfine interaction with the defect could help to identify and monitor the involved intermediate states after the recombination.

Another possible explanation could be a mw-induced excitation of electrons from the OV$^-$ state into the conduction band.  Since the OV$^-$ is charged, the scattering cross section with conduction band electrons accelerated by the mw electric field is expected to be much larger than that of the neutral $^{31}$P donor or the SL1 center. In addition, the potential of the nearby ionized donor $^{31}$P$^+$ would possibly lower the energy barrier for the emission of an electron. \cite{milnes_deep_1973} However, this interpretation is at variance with the observed resonant enhancement of current after a single mw pulse, since the OV$^-$ form only after the spin-dependent transition which takes place on a timescale of 4~$\mu$s, much longer than the length of the mw pulse.

Independent of its microscopic interpretation, the observed excitation of carriers without illumination provides an interesting alternative for pulsed EDMR detection in these samples. It does not suffer from light-induced background noise and the current transients are recorded with higher sensitivity, significantly improving the signal-to-noise ratio compared to measurements with continuous illumination.

\section{Recombination dynamics}\label{secRecDyn}

\begin{figure}%
\includegraphics[width=\columnwidth]{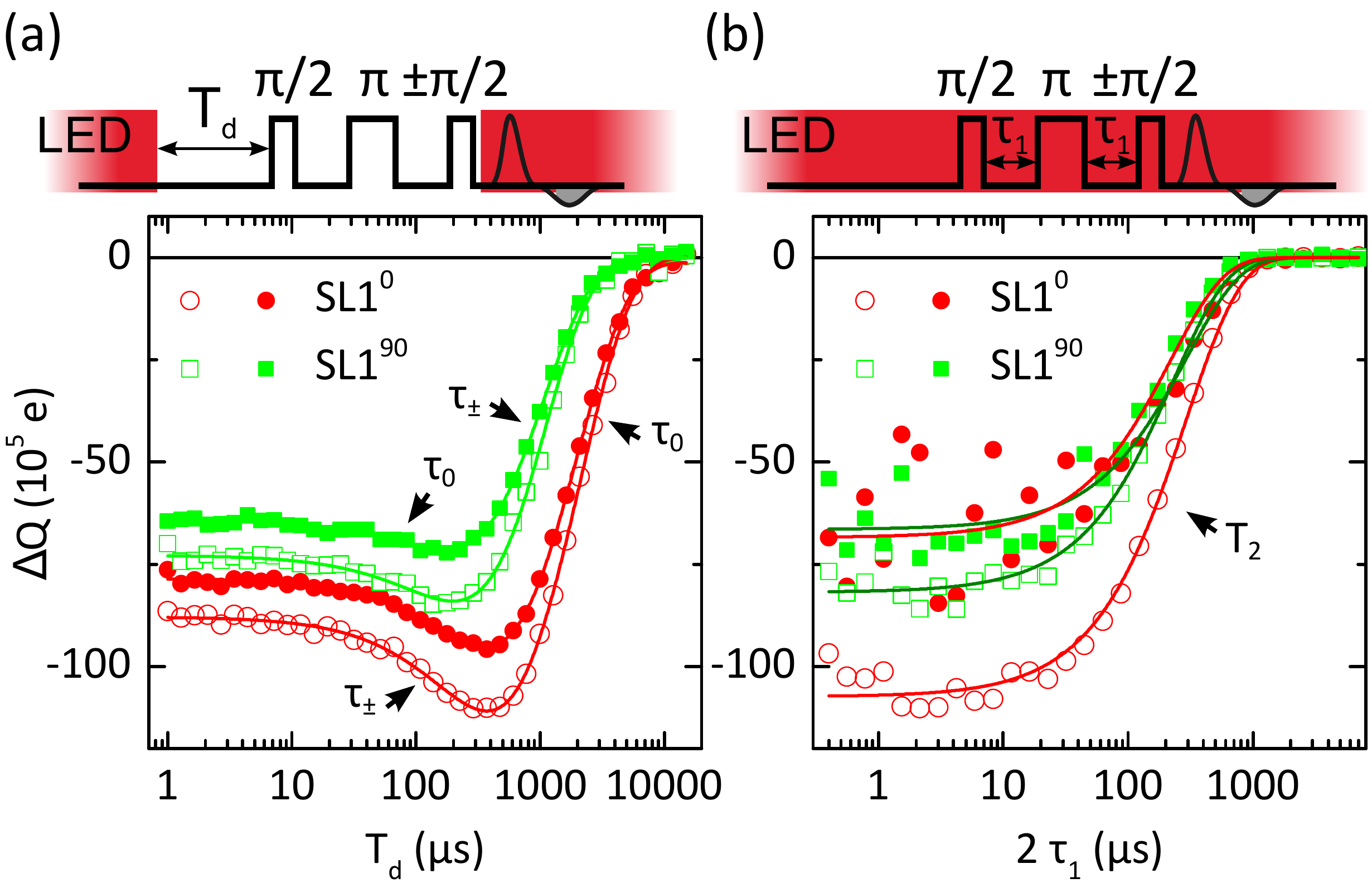}%
\caption{Determination of timeconstants for the SL1-only recombination process. Open and closed symbols represent the low-field and high-field SL1 transitions, resp., for the given orientation. (a) Illumination delay experiment yielding the lifetimes $\tau_{\pm}$ and $\tau_0$ of the different triplet states. (b) Echo decay for the determination of the coherence time. Solid lines represent fits with exponential functions.}%
\label{figTimeConstSL1}%
\end{figure}

\begin{table*}%
\begin{tabular}{lcccccl}
transition &$0\leftrightarrow -1$, SL1$ ^{0}$  & $0\leftrightarrow 1$, SL1$^{0}$ & $1\leftrightarrow 0$, SL1$^{90}$ & $-1\leftrightarrow 0$, SL1$^{90}$&\\
\hline
magnetic field (mT) & 312.4 & 382 & 329.7  & 362.4&this work\\
\hline
\hline
$\tau_0$ ($\mu$s)  & $2024 \pm 35$& $1992\pm40$& $222\pm16$  & $196 \pm 18$&this work\\
 & $1970\pm4$ & $2000\pm 4$ & $205\pm 1$ & $200\pm 1$&Ref.~\onlinecite{akhtar_coherent_2012}\\
\hline
$\tau_{\pm}$ ($\mu$s) & $283 \pm 12$& $287\pm 15$  & $947\pm30$  & $955\pm50$&this work\\
 & $330 \pm 2$ & $280\pm 2$ & $987\pm 2$ & $960\pm2$&Ref.~\onlinecite{akhtar_coherent_2012}\\
\hline
$T_2$ ($\mu$s)  & $293\pm 11$ & $296 \pm 60$ & $230\pm 10$  & $218\pm25$&this work\\
 & & $240\pm 4$ & & &Ref.~\onlinecite{akhtar_rabi_2012}
\end{tabular}
\caption{Time constants for SL1-only recombination processes. The magnetic field values are given for $f=9.74$ GHz [cf.~Fig.~\ref{pEDMRSpec}(b)].}
\label{tabSL1tc}
\end{table*}

Pulsed EMDR, especially when combined with pulsed illumination, provides powerful tools for the study of the lifetimes involved in spin-dependent processes. \cite{hoehne_time_2013} In particular, the lifetimes of the involved states can be measured with high precision, which is essential for the design of complex pulse sequences, e.g., ENDOR experiments.\cite{hoehne_electrical_2011, dreher_nuclear_2012, hoehne_spin_2013} For both of the two processes described in this work, an EDMR signal is observed because of the mw-induced transition from a longer-lived to shorter-lived state.

We first consider the time constants involved in the spin-dependent recombination process via the SL1 center only, which are the lifetimes $\tau_0$ and $\tau_{\pm}$ of the SL1 states $T_0$ and $T_\pm$, respectively (c.f.~Fig.~\ref{figRec_process}). To access these time constants, we use an illumination delay experiment as sketched in Fig.~\ref{figTimeConstSL1}(a). After illuminating the sample for several milliseconds the LED is switched off and, after a delay $T_{\mathrm{d}}$, an echo sequence is applied. The transient change in photocurrent is recorded under illumination and the integration interval is chosen such that only the slower process is monitored [$\Delta = 450-950~\mu$s, gray shaded area in Fig.~\ref{figTimeConstSL1}(a)]. 
The results are shown in Fig.~\ref{figTimeConstSL1}(a), where the echo amplitude $\Delta Q$ is shown as a function of $T_{\mathrm{d}}$ for the transitions of the two detectable SL1 orientations, SL1$^0$ and SL1$^{90}$. For all transitions, the absolute echo amplitude is first enhanced for $T_\mathrm{d}$ around 200 to 400 $\mu$s before decaying to zero for longer $T_{\mathrm{d}}$. Without illumination, no more defects are excited into the triplet state and the relaxation of the short-lived triplet states ($T_{\pm}$ for SL1$^0$, $T_0$ for SL1$^{90}$) leads to a higher polarization of the SL1 and a rise of the echo amplitude. The signal then decays with a time constant corresponding to the relaxation of the long-lived triplet states to the singlet ground state. The time constants were extracted by exponential fitting of the data [solid lines in Fig.~\ref{figTimeConstSL1}(a)] and are given in Tab. \ref{tabSL1tc}. They are in very good agreement with previous measurements of the lifetime of the SL1 triplet states by EPR. \cite{akhtar_coherent_2012}

Furthermore, the spin coherence time $T_2$ was determined by measuring the echo amplitude $\Delta Q$ as a function of the free evolution time $2\tau_1$ as shown in Fig.~\ref{figTimeConstSL1}(b). A strong oscillation of the echo amplitude is observed due to electron spin echo envelope modulation (ESEEM) effects caused by the hyperfine interactions with $^{29}$Si nuclear spins. \cite{akhtar_rabi_2012} The modulation depth depends on the strength and anisotropy of the hyperfine interactions of the involved SL1 states and is therefore different for the four transitions shown in Fig.~\ref{figTimeConstSL1}(b). The values for $T_2$ as obtained by fitting with exponential functions are also given in Tab.~\ref{tabSL1tc}. For all transitions they are similar to the lifetimes of the shorter-lived state ($\tau_{\pm}$ for SL1$^0$, $\tau_0$ for SL1$^{90}$), which suggests that the observed transverse relaxation time $T_2$ is in fact limited by these lifetimes. The observed $T_2\approx 220~\mu$s is comparatively long for a spin coherence time in EDMR, where typically values of less than $10~\mu$s are observed at similar experimental conditions.\cite{huebl_spin_2008, paik_t_1_2010} Only in high fields ($\approx8.6$ T), values of $T_2\approx160~\mu$s have been observed.\cite{morley_long-lived_2008} For the SL1 center, $T_2= 240~\mu$s has been measured in pulsed EPR experiments, which is in very good agreement with our results. $T_2$ has been reported to be strongly reduced under illumination.\cite{akhtar_rabi_2012} This is in contrast to the results of our experiments where $T_2$ is not affected by the continuous illumination, probably due to the comparatively low power of the LED ($P_{\mathrm{LED}}\approx 60~\mathrm{mW}/\mathrm{cm}^2$) used in this work.

\begin{figure}%
\includegraphics[width=\columnwidth]{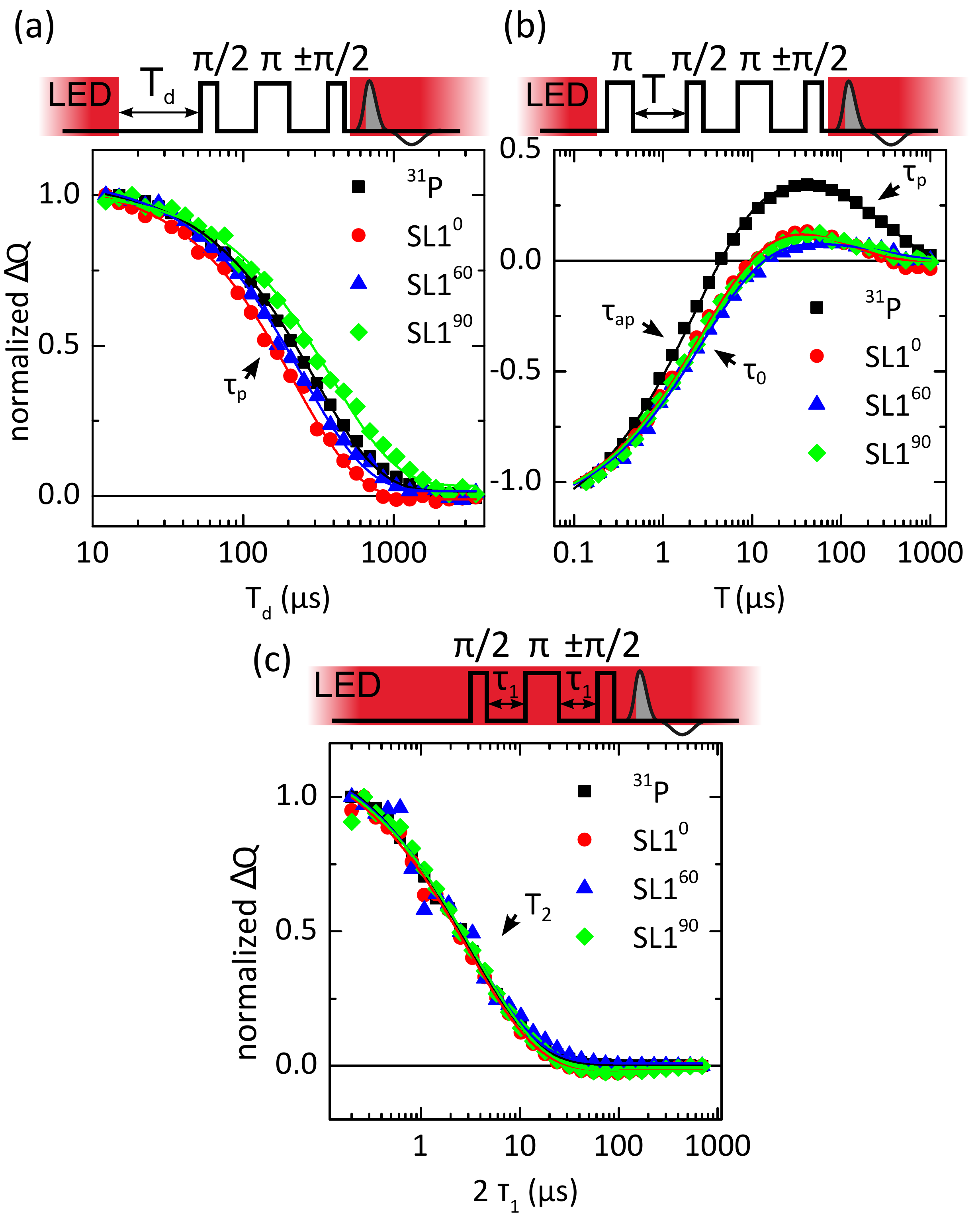}%
\caption{Determination of parallel and antiparallel lifetimes for the $^{31}$P--SL1 pair process. Solid lines represent fits with an exponential decay function in (a) and with a stretched exponential function (exponent $n=0.6$) for the faster decay in (b). The resulting effective time constants are given in Tab. \ref{tabPairtc}. (c) Echo decay experiment yielding the coherence time $T_2$.}%
\label{figTimeConstPair}%
\end{figure}

Until now, we have studied the dynamics of the SL1-only recombination. For the determination of the lifetimes of the $^{31}$P--SL1 spin pairs, we repeat the illumination delay experiment with an integration interval chosen such that only the faster pair process is monitored ($\Delta = 1-5~\mu$s). Though illuminating again after the pulse sequence is not necessary for the detection of the signal connected this process (c.f.~Sect.~\ref{secEnhancement}), the pulse sequence was not altered to ensure consistency with the measurements shown in Fig.~\ref{figTimeConstSL1}(a).
Figure \ref{figTimeConstPair} (a) shows the integrated charge $\Delta Q$ as a function of $T_{\mathrm{d}}$ for the high-field $^{31}$P resonance, as well as one transition for each of the three SL1 orientations.
In this case, a single decay is observed. Fitting it with an exponential (solid lines) yields the lifetime of the longer-lived (parallel) state $\tau_\mathrm{p}$ listed in Tab.~\ref{tabPairtc}. We observe different values for the three SL1 orientations. While for SL1$^0$ the lifetimes of the parallel pairs might be given by the lifetimes $\tau_{\pm}$ of the $T_{\pm}$ states of SL1, this does not seem to be the limiting mechanism for the SL1$^{90}$ orientation, where $\tau_{\pm}$ is significantly longer than $\tau_{\mathrm{p}}$.

The expected rise of the echo amplitude with a time constant given by the shorter lifetimes $\tau_{\mathrm{ap}}$ and $\tau_0$ happens on a timescale similar to the fall time of the illumination pulse ($\approx 3~\mu$s), which is why the illumination delay sequence in Fig.~\ref{figTimeConstPair}(a) is not suitable for their precise determination in the case of the pair process. 
Therefore, an inversion recovery experiment \cite{paik_t_1_2010} was performed as shown in Fig.~\ref{figTimeConstPair}(b). When using pulsed illumination, the inversion recovery signal $\Delta Q$ decays to zero and the lifetime $\tau_{\mathrm{ap}}$ can be extracted without having to consider the generation of new pairs.\cite{hoehne_time_2013} Since $\tau_{\mathrm{ap}}$ most likely depends on the distance between donor and defect and therefore a distribution of timeconstants is expected, the decay of the inverted amplitude is fitted with a stretched exponential function ($n=0.6$) as also done to quantitatively describe the $^{31}$P--P$_{\mathrm{b}0}$ recombination dynamics.\cite{hoehne_time_2013} The corresponding effective lifetimes are given in Tab.~\ref{tabPairtc}. Furthermore, a second decay with the parallel lifetime $\tau_{\mathrm{p}}$ is observed due to the imperfection of the inversion pulse. The two measurements of $\tau_\mathrm{p}$ [Fig.~\ref{figTimeConstPair}(a) and \ref{figTimeConstPair}(b)] are in very good agreement.

In addition, the spin coherence time $T_2$ was measured also for the pair process [Fig.~\ref{figTimeConstPair}(c)]. The observed decay can be fitted with a stretched exponential ($n=0.6$) and is very similar to the faster decay in the inversion recovery experiment [$\tau_{\mathrm{ap}}$ in Fig.~\ref{figTimeConstPair}(b)]. It is therefore likely that the decay is limited by $\tau_{\mathrm{ap}}$.

\begin{table*}%
\begin{tabular}{lcccc}
transition & $-1\leftrightarrow 0$, SL1$^{0}$& $1\leftrightarrow 0$, SL1$^{90}$ & SL1$^{60}$ & $^{31}$P\\
\hline
magnetic field (mT)& 312.4 & 329.7 & 340 & 346.4\\
\hline
\hline
$\tau_{\mathrm{p}}$ ($\mu$s) & $213\pm 5$ & $420 \pm13$ & $257\pm 6$ & $293\pm4$\\
$\tau_0$ ($\mu$s)  & $4.1 \pm 0.3$ & $4.1 \pm 0.3$ & $4.1 \pm 0.3$ &\\
$\tau_{\mathrm{ap}}$ ($\mu$s) & & & &$4.2 \pm 0.2$\\

\hline
$T_2$ ($\mu$s) & $4.4\pm0.4$ & $4.8 \pm 0.4$ & $4.5 \pm 0.4$ & $4.3\pm 0.4$ 
\end{tabular}
\caption{Time constants for the $^{31}$P--SL1 pair processes. For values obtained from fitting with a streched exponential ($\tau_0$, $\tau_{\mathrm{ap}}$ and $T_2$), the effective decay time $\tau_{\mathrm{eff}}=\tau/n\cdot\Gamma(1/n)$ is given. For details see Ref. \onlinecite{berberan-santos_mathematical_2005}. The magnetic field values are given for $f=9.74$ GHz [cf.~Fig.~\ref{pEDMRSpec}(b)].}
\label{tabPairtc}
\end{table*}

\section{Zero-field EDMR}
The strength of the spin orbit interaction and the resulting lifetimes of the triplet states depend on the orientation of the spin system's quantization axis within the defect geometry.\cite{vlasenko_epr_1984,vlasenko_electron_1995} Therefore it can be described by three time constants $\tau_x$, $\tau_y$ and $\tau_z$, which are connected to the defect's three principle axes. As shown below, we can derive these values from the lifetimes measured for the high-field states $\tau_{\pm}$ and $\tau_0$ for different orientations of the defect with respect to $B_0$. It is also possible to obtain the lifetimes $\tau_x$, $\tau_y$ and $\tau_z$ directly by measuring them in an experiment without external magnetic field.
Indeed, pulsed zero-field EPR has been used to study the generation and relaxation mechanisms of the SL1 triplet states.\cite{frens_dynamics_1993}

While spin-dependent recombination at zero field is used quite commonly to determine the zero-field splittings of $S=1$ systems by optically detected magnetic resonance (ODMR),\cite{el-sayed_double_1975, salib_zero-field_1984} EDMR has so far only be performed at small but finite magnetic fields,\cite{brandt_electrically_1998, morishita_electrical_2009, baker_robust_2012, cochrane_zero-field_2012} where it has proven to be a very valuable spectroscopic tool, since the signal intensity does not depend on thermal polarization but on the relative orientation of two partners in a pair process.\cite{kaplan_explanation_1978} In our sample, the resonance of the $S=1/2$ phosphorus donor can be observed for magnetic fields as low as $1$ mT. At even lower magnetic fields, the signal vanishes because parallel and antiparallel spin configurations of a $^{31}\mathrm{P}$--SL1 pair spin pair can only form for the SL1 high-field states $T_{\pm}$ (c.f.~Fig.~\ref{figELDORstates}). However, the signal for the SL1-only recombination process is observed independently of the magnetic field, allowing the realization of a true zero-field cw and pulsed EDMR.

\begin{figure}%
\includegraphics[width=\columnwidth]{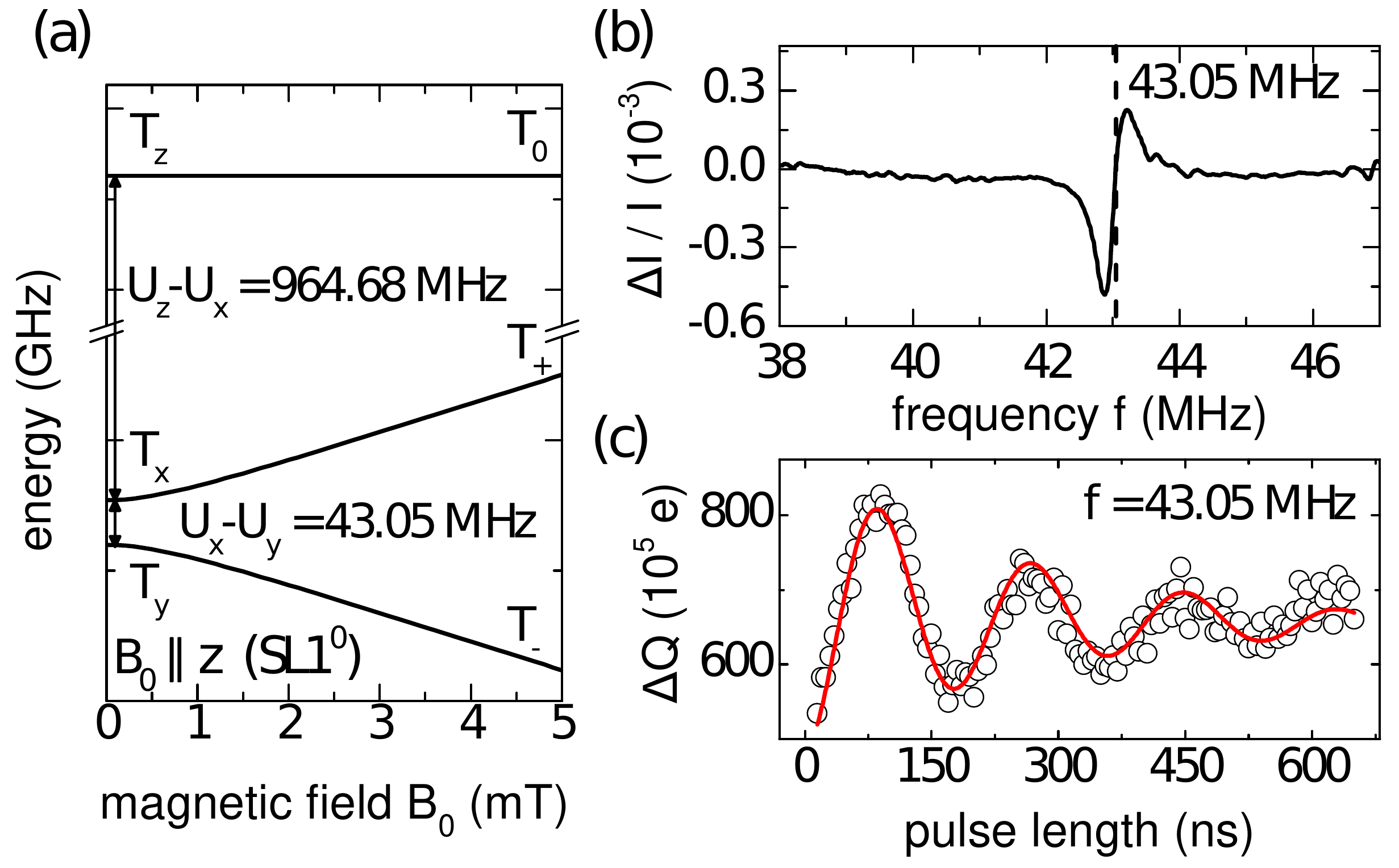}%
\caption{(a) SL1 energy level diagram for a magnetic field along the $z$-axis. $T_x$, $T_y$ and $T_z$ indicate the corresponding spin states at zero external magnetic field, the labels $T_+$, $T_0$ and $T_-$ show the spin states  in the high-field limit. (b) Continuous wave zero-field EDMR spectrum recorded as a function of the radio frequency $f$ for $B_0=0$. (c) Rabi oscillations for the $T_x\leftrightarrow T_y$ transition observed in pulsed zero-field EDMR.}%
\label{figzerofield}%
\end{figure}

For $B_0=0$, the SL1 center is described by the three zero-field states $T_\mathrm{x}$, $T_y$ and $T_z$ with the energies
\begin{align*}
	U_\mathrm{x}&= 1/3\cdot D-E\text{ ,}\\ U_y&=1/3\cdot D +E \text{ and}\\
	U_z&= -2/3\cdot D\text{ ,}
\end{align*}
where $D=-986.2$ MHz and $E= 21.525$ MHz.\cite{brower_electron_1971} The eigenstates of the system at zero magnetic field are given in the defect's coordinate system introduced in Fig.~\ref{figSL1}(a).
The high-field triplet states $T_-$, $T_0$ and $T_+$, which describe the system for $\mu_B\vec{B}_0 g\vec{S}\gg h/2\cdot\vec{S}D\vec{S}$, are superpositions of the zero-field states. Considering the orientations SL1$^0$ and SL1$^{90}$, the corresponding states $T_{j}^0$ and $T_{j}^{90}$ ($j\in\{+,0,-\}$) are given by
\begin{align*}
	T_+^0&=\tfrac{1}{\sqrt{2}}(T_{\mathrm{x}}+iT_{y})\text{ ,} & T_+^{90}&=\tfrac{1}{\sqrt{2}}(T_z+iT_{\mathrm{x}})\text{ ,}\\
	T_0^0&= T_z\text{ ,} & T_0^{90}&= T_{y}\text{ ,}\\
	T_-^0&= \tfrac{1}{\sqrt{2}}(T_{\mathrm{x}}-iT_{y})\text{ ,}& T_-^{90} &= \tfrac{1}{\sqrt{2}}(T_z-iT_{\mathrm{x}})\text{ .}	  
\end{align*}
The transition rate $R$ to the singlet ground state $S_0$ is proportional to the square of the matrix element of the spin orbit interaction Hamiltonian $\mathcal{H}_{SO}$ regarding the two states
\begin{align}
	R_{j}=\frac{1}{\tau_{j}}\propto \left|\langle S_0\mid \mathcal{H}_{\mathrm{SO}} \mid T_j\rangle\right|^2\text{.}
\end{align}
The lifetimes $\tau_j$ of the different triplet states are therefore connected by
\begin{align}\label{eqRecRates}
	\tau_0^{0}&=\tau_z\text{ ,} & \tau_+^{0}=\tau_-^{0}&=\left(\frac{1}{2\tau_{\mathrm{x}}}+\frac{1}{2\tau_{y}}\right)^{-1}\text{ ,}\\
	\tau_0^{90}&=\tau_y\text{ ,} & \tau_+^{90}=\tau_-^{90}&=\left(\frac{1}{2 \tau_{z}}+\frac{1}{2\tau_{\mathrm{x}}}\right)^{-1}\text{,}
\end{align}
where $\tau_{j}^{0}$ and $\tau_{j}^{90}$ denote the high-field lifetimes for an external magnetic field along the $z$- and $y$-axis, respectively.

\begin{table}%
\begin{tabular}{lcccc}
 & $\tau_x$ ($\mu$s) & $\tau_y$ ($\mu$s) & $\tau_z$ ($\mu$s)\\
\hline
\hline
high field & $535\pm 110$ & $209\pm25$ & $2008\pm40$& this work\\
zero field & $995\pm35$ & $270\pm5$ &  & this work\\
zero field & $790\pm40$ & $220\pm11$ & $1410\pm70$ & Ref.~\onlinecite{frens_dynamics_1993}
\end{tabular}
\caption{Triplet lifetimes as determined by pulsed EDMR at $B_0\approx 350$ mT and at vanishing magnetic field. Values taken from Ref. \onlinecite{frens_dynamics_1993} were determined in pulsed zero-field EPR experiments and are shown for comparison.}
\label{tabZfTc}
\end{table}
The SL1 energy level diagram for small external magnetic fields in $z$-direction is shown in Fig.~\ref{figzerofield}(a), where the traces are labeled with the zero-field states on the left and with the high-field states on the right side of the plot. Since the maximum radio frequency (rf) is limited in the experimental setup used, only the transition $T_{x}\leftrightarrow T_{y}$ could be studied in this work. We first detect this transition by zero-field cw EDMR. Figure \ref{figzerofield} (b) shows the change in photocurrent $\Delta I / I$ under continuous illumination as a function of the radio frequency $f$. The spectrum was recorded under continuous wave rf irradiation via the ENDOR coils of the resonator with frequency modulation and lock-in detection. The external magnetic field was tuned to zero using a Hall magnetometer connected to a bipolar power supply. A pronounced peak is observed at the expected frequency of $f_{xy}=43.05$ MHz [cf.~Fig.~\ref{figzerofield}(a)]. The coherent control of the SL1 spin system at zero field via pulsed EDMR is shown in Fig.~\ref{figzerofield}(c), where Rabi oscillation were recorded using pulsed $43.05$ MHz irradiation amplified by a $300$ W power amplifier.

The lifetime $\tau_x$ of the long-lived state for this transition was determined in an illumination delay experiment such as shown in Fig.~\ref{figTimeConstPair}(a) but performed at zero field. Fitting with a single exponential function yields $995\pm35~\mu$s. The lifetime of the short-lived state, assigned to $\tau_y$, was measured in an inversion recovery experiment [cf.~Fig.~\ref{figTimeConstPair}(b)] yielding $\tau_y=270\pm5~\mu$s. Both values are listed in Tab.~\ref{tabZfTc} in comparison to the expected lifetimes calculated using (\ref{eqRecRates}) and the high-field lifetimes as given in Tab. \ref{tabSL1tc}. The two accessible lifetimes are significantly longer for $B_0=0$. This could possibly be explained by the mixing of singlet and triplet states caused by the magnetic field. Even though the SL1 is still well described as an effective $S=1$ system at $B_0\approx 350$ mT, even a small singlet content can lead to a notable reduction in the triplet lifetime.

Both SL1 transitions at zero field are so-called clock transitions or optimal working points because their transition frequences are to first order magnetic field-independent ($\partial f/\partial B_0 =0$ for $B_0=0$).\cite{mohammady_analysis_2012, wolfowicz_atomic_2013} Since the spin state is insensitive to magnetic field noise at these transitions, longer coherence times can be observed. For the $T_{x}\leftrightarrow T_{y}$ transition, we observed a value of $T_2= 180\pm6$ $\mu$s in an echo decay measurement [cf.~Fig.~\ref{figTimeConstPair}(c)]. This is shorter than the lifetimes of the involved triplet states $\tau_x$ and $\tau_y$ and suggests that in this case the coherence time is no longer limited by these lifetimes and is in fact shorter than the coherence time at high fields. However, to completely exclude that $T_2$ is limited by $\tau_y$, the experiment should be repeated for the transition $T_{x}\leftrightarrow T_{z}$, where otherwise a higher value of up to 1 ms could be expected due to the longer lifetime of the involved triplet states. This, however, would require rf frequencies at $964.68$ MHz, which could not be delivered by the pulsed ENDOR resonator used.

\section{Summary and Outlook}
In this paper we presented experimental evidence for two different spin-dependent processes in phosphorus-doped silicon after irradiation with high energy photons: 
(i) the spin-dependent relaxation of the SL1 triplet to the singlet ground state leading to a resonant enhancement in charge carrier recombination and (ii) a donor-acceptor-like recombination process where a $^{31}$P electron is transferred to the SL1 depending on the relative orientation of their spin states. In the latter case, a resonant increase of the conductivity of the sample is observed. As a possible explanation for the microscopic origin of this increase we suggest the Auger excitation of an electron of the OV$^{-\ast}$ state to the conduction band. Independently of this mechanism, the discussed pair recombination process provides a sensitive read-out mechanism for the electrical detection of $^{31}$P donors in bulk silicon, a method that might also be applicable to other group-V donors like arsenic, antimony or bismuth. From a more general point of view, we showed that a spin pair involving an $S=1$ and an $S=1/2$ system can lead to Pauli blockade and a spin-dependent charge transfer. This could allow to realize an electrical read-out of the other $S=1$ centers, in particular the negative charge state of the NV center in diamond.

In addition, the recombination dynamics were studied for both processes. In particular, the lifetimes of the involved spin states were determined in pulsed EDMR experiments. The values obtained for the SL1-only relaxation process are in very good agreement with previous measurement with pulsed EPR. The recombination times found for the $^{31}$P--SL1 pair process are found to be comparable to the corresponding times for $^{31}$P--P$_{\mathrm{b}0}$ recombination at the Si/SiO$_2$ interface.\cite{hoehne_time_2013} Finally, zero-field EDMR measurements for the SL1 were presented and the transition between two of the three zero-field states was studied using zero-field pulsed EDMR. The time constants obtained at zero field are longer than the high-field values and again in good agreement with EPR measurements. At zero field, the coherence time $T_2$ is found to be slightly reduced when compared to high-field values.

In summary, the work presented here shows that pulsed EDMR can also be successfully applied to spin systems more complicated than the $S=1/2$ spin pairs studied until now, providing a wealth of information on recombination pathways and dynamics.

\section*{Acknowledgments}
This work was funded by DFG (Grant No. Br 1585/5 and Br 1585/7), the JST-DFG Strategic Cooperative Program on Nanoelectronics, the Core-to-Core Program by JSPS, FIRST, and a Grant-in-Aid for Scientific Research and Project for Developing Innovation Systems by MEXT.

\bibliography{SL1}

\end{document}